\documentclass[trackchanges,twocolumn,astrosymb,twocolappendix]{aastex701}

\usepackage{gensymb}

\shorttitle{COSI sGRB Detection Predictions}
\shortauthors{Neights et al.}

\submitjournal{ApJ}

\graphicspath{{./}}

\begin{document}

\title{Predictions for Short Gamma-Ray Burst Detections by COSI}

\author[0009-0005-0762-4507]{Eliza~Neights}
\affiliation{Department of Physics, The George Washington University, Washington, DC, USA}
\affiliation{Astrophysics Science Division, NASA Goddard Space Flight Center, Greenbelt, MD, USA}
\email{eliza.neights@gmail.com}

\author[0000-0002-2942-3379]{Eric~Burns}
\affiliation{Department of Physics \& Astronomy, Louisiana State University, Baton Rouge, LA, USA}
\email{ericburns@lsu.edu}

\author[0000-0002-4535-5329]{Nicolò Parmiggiani}
\affiliation{INAF - Osservatorio di Astrofisica e Scienza dello spazio di Bologna (OAS), Bologna, Italy}
\email{nicolo.parmiggiani@ inaf.it}

\author[0000-0002-3833-1054]{Yong Sheng}
\affiliation{Department of Physics and Astronomy, Clemson University, Kinard Lab of Physics, Clemson, SC, USA}
\email{sheng2@g.clemson.edu}

\author[0000-0002-2664-8804]{Savitri Gallego}
\affiliation{Institut für Physik \& Exzellenzcluster PRISMA+, Johannes Gutenberg-Universität Mainz, Mainz, Germany}
\email{sgallego@uni-mainz.de}

\author[0000-0002-3904-4487]{Aldo Aguilar-Nadalini}
\affiliation{Space Sciences Laboratory, University of California Berkeley, Berkeley, CA, USA}
\email{aldo.aguilar@berkeley.edu}

\author[0009-0006-8598-728X]{Aaron C. Trigg}
\affiliation{NASA Postdoctoral Program Fellow, NASA Marshall Space Flight Center, Huntsville, AL, USA}
\email{aaron.c.trigg@nasa.gov}

\author[0000-0001-5780-8770]{Sylvain Guiriec}
\affiliation{Department of Physics, The George Washington University, Washington, DC, USA}
\affiliation{Astrophysics Science Division, NASA Goddard Space Flight Center, Greenbelt, MD, USA}
\email{sguiriec@gwu.edu}

\author[0009-0008-7888-7584]{Koothodil Abhijith Augustine}
\affiliation{Institute of Astronomy, National Tsing Hua University, Hsinchu, Taiwan}
\email{abhijithaugustine007@gmail.com}

\author[0000-0003-4433-1365]{{Matthew G.} {Baring}}
\affiliation{Department of Physics and Astronomy, Rice University, Houston, TX, USA}
\email{baring@rice.edu}

\author[0000-0002-6439-1472]{Anna L. Butterworth}
\affiliation{Space Sciences Laboratory, University of California Berkeley, Berkeley, CA, USA}
\email{butterworth@berkeley.edu}

\author[0000-0002-5617-3117]{Hsiang-Kuang Chang}
\affiliation{Institute of Astronomy, National Tsing Hua University, Hsinchu, Taiwan}
\email{hkchang@mx.nthu.edu.tw}

\author[0009-0002-2845-2343]{Che-Yen Chu}
\affiliation{Department of Physics, National Changhua University of Education, Changhua, Taiwan}
\email{cychu@gapp.nthu.edu.tw}

\author[0000-0002-7574-1298]{Niccol\`{o} Di Lalla}
\affiliation{W. W. Hansen Experimental Physics Laboratory, Stanford University, Stanford, CA, USA}
\affiliation{Kavli Institute for Particle Astrophysics and Cosmology, SLAC National Accelerator Laboratory \& Stanford University, Stanford, CA, USA}
\affiliation{Department of Physics, Stanford University, Stanford, CA, USA}
\email{niccolo.dilalla@stanford.edu}

\author[0000-0003-2624-0056]{Christopher Fryer}
\affiliation{Center for Nonlinear Studies, Los Alamos National Laboratory, Los Alamos, NM, USA}
\affiliation{Department of Physics, The George Washington University, Washington, DC, USA}
\email{fryer@lanl.gov}

\author[0000-0002-2752-9524]{Hadar Lazar}
\affiliation{Space Sciences Laboratory, University of California Berkeley, Berkeley, CA, USA}
\email{lazar.hadar@gmail.com}

\author[0000-0002-2471-8696]{Israel Martinez-Castellanos}  
\affiliation{University of Maryland, College Park, MD, USA} 
\affiliation{Astrophysics Science Division, NASA Goddard Space Flight Center, Greenbelt, MD, USA}   
\email{imc@umd.edu}

\author[0000-0003-2318-7763]{Marion Sudvarg}
\affiliation{Missouri University of Science and Technology, Rolla, MO, USA}
\email{msudvarg@mst.edu}

\author[0000-0002-2149-9846]{P\'eter Veres}
\affiliation{Department of Space Science, University of Alabama in Huntsville, Huntsville, AL, USA}
\affiliation{Center for Space Plasma and Aeronomic Research, University of Alabama in Huntsville, Huntsville, AL, USA}
\email{peter.veres@uah.edu}

\author[0000-0001-9567-4224]{Steven E. Boggs}
\affiliation{Department of Astronomy and Astrophysics, University of California San Diego, La Jolla, CA, USA}
\affiliation{Space Sciences Laboratory, University of California Berkeley, Berkeley, CA, USA}
\email{seboggs@ucsd.edu}

\author[0000-0002-8028-0991]{Dieter H. Hartmann}
\affiliation{Department of Physics and Astronomy, Clemson University, Kinard Lab of Physics, Clemson, SC, USA}
\email{hdieter@g.clemson.edu}

\author[0000-0001-6677-914X]{Carolyn A. Kierans}
\affiliation{Astrophysics Science Division, NASA Goddard Space Flight Center, Greenbelt, MD, USA}
\email{carolyn.a.kierans@nasa.gov}

\author[0000-0001-5506-9855]{John A. Tomsick}
\affiliation{Space Sciences Laboratory, University of California Berkeley, Berkeley, CA, USA}
\email{jtomsick@berkeley.edu}

\author[0000-0001-9067-3150]{Andreas Zoglauer}
\affiliation{Space Sciences Laboratory, University of California Berkeley, Berkeley, CA, USA}
\email{zoglauer@berkeley.edu}

\begin{abstract}

The Compton Spectrometer and Imager (COSI) is a soft gamma-ray telescope scheduled to launch in 2027. COSI will trigger onboard the spacecraft in response to short-duration transients, including gamma-ray bursts, and it will promptly disseminate alerts which include localizations to the community. This will enable rapid follow-up observations of the afterglow and associations with transients identified by other instruments. By simulating COSI observations of short~gamma-ray bursts (sGRBs) and processing them with the trigger algorithm which will run onboard the spacecraft, we estimate the sensitivity of COSI to sGRBs. We then convolve this with a simulated set of sGRBs representing the intrinsic population in the universe to predict the rate that will be detected by COSI. We find that COSI is expected to trigger on $\sim$$47-89$~sGRBs per year with $\sim$$7-10$~localized to within~$2.5\degree$. With the gravitational wave sensitivity anticipated in the O5 run, we predict $0-3$~joint detections with gravitational waves.

\end{abstract}

\keywords{\uat{Gamma-ray bursts}{629} --- \uat{High Energy astrophysics}{739} --- \uat{Gravitational waves}{678} --- \uat{Gamma-ray telescopes}{634}} 

\section{Introduction} \label{sec:intro}

Gamma-ray bursts (GRBs) are the most luminous explosions in the universe. Their prompt gamma-ray emission is produced by an ultrarelativistic jet powered by a stellar-mass compact object. A broadband afterglow follows when the jet interacts with the medium surrounding the burst. Despite over half a century of observations, the progenitors and physics of the GRB jet remain poorly understood and it is unknown what all of the progenitors are. 

Short GRBs (sGRBs) have a prompt gamma-ray duration $T_{90}<2$~s \citep{1984Natur.308..434N,1993ApJ...413L.101K}. Most sGRBs are thought to originate from neutron star mergers, with a small fraction known to be the result of magnetar giant flares in nearby galaxies \citep{2006ApJ...652..507O,2008ApJ...680..545M,2021ApJ...907L..28B,2026A&A...710A.139T}. Neutron star mergers can refer to either the merger of two neutron stars or a neutron star and a black hole, and their production of sGRBs is evidenced by kilonova detections associated with sGRBs \citep{2013Natur.500..547T,2013ApJ...774L..23B,2015ApJ...811L..22J,2016NatCo...712898J} and sGRB locations in the outskirt or outside of their host galaxies \citep{1999MNRAS.305..763B,1999ApJ...526..152F,2013ApJ...776...18F,2014MNRAS.437.1495T}. They were confirmed as a sGRB progenitor by the joint detection of GRB~170817A and GW170817 \citep{2017ApJ...848L..13A,2017ApJ...848L..14G} and the associated kilonova detection \citep{2017Natur.551...71T,2017Sci...358.1565E,2017Natur.551...64A}. Additional joint sGRB and gravitational wave (GW) detections could enable a greater understanding of sGRB progenitors, jet structure and composition, and gamma-ray emission mechanisms, as well as the fraction of heavy elements forged in neutron star mergers and fundamental physics \citep{2018PhRvD..97h3013S,2018MNRAS.478..733L,2018MNRAS.479..588G,2018MNRAS.481.1597G,2019ApJ...877L..40G}.

The Compton Spectrometer and Imager (COSI) is a soft gamma-ray space telescope with a planned launch in 2027 which will provide high-resolution spectroscopy and polarimetry of GRBs in the 200~keV~$-$~5~MeV range \citep{2024icrc.confE.745T}. COSI will trigger on GRBs and other fast transient events onboard and rapidly alert the community. Localizations of GRBs performed on the ground will be shared with the community within an hour of detection to enable multi-wavelength follow-up. The onboard trigger algorithm is optimized for the detection of sGRBs in order to maximize the potential for joint sGRB and GW events. The COSI requirement is to localize sGRBs with a signal-to-noise ratio (SNR)~$>$~6 to within 2.5$\degree$~uncertainty at 90\%~confidence.

The sGRB positions determined by COSI may be followed up by optical and infrared telescopes, enabling rapid follow-up by multi-wavelength instruments and associations with externally-detected transients such as kilonovae \citep{2023arXiv230804485B}. Joint sGRB and kilonova detections may be possible through the Vera Rubin Observatory Legacy Survey of Space and Time \citep{2020arXiv200111067B} and rapid target of opportunity observations by the Vera Rubin Observatory. Additionally, the O5~run of the Laser Interferometer Gravitational-Wave Observatory \citep[LIGO;][]{2015CQGra..32g4001L}, Virgo interferometer \citep{2015CQGra..32b4001A}, and Kamioka Gravitational Wave Detector \citep[KAGRA;][]{2021PTEP.2021eA103A} scheduled to begin in 2029 will feature upgraded detectors with increased sensitivity. This enables possible joint detections between GWs observed by LIGO-Virgo-KAGRA and sGRBs detected by COSI, primarily after its two-year prime mission duration.

In this paper, we report predictions for the rate of sGRBs that will be detected and well-localized by COSI, as well as expectations for coincident GW detections. In Section \ref{sec:cosi}, we introduce COSI, its onboard trigger algorithm, and transient analysis pipelines. We describe the simulated COSI observations of sGRBs used to determine the trigger efficiency of COSI in Section \ref{sec:simulations}, and we discuss the predicted sGRB detection rate and joint COSI GRB-GW rate in Section \ref{sec:sgrb-rate}.

\section{The Compton Spectrometer and Imager} \label{sec:cosi}

\begin{figure*}[ht!]
    \plotone{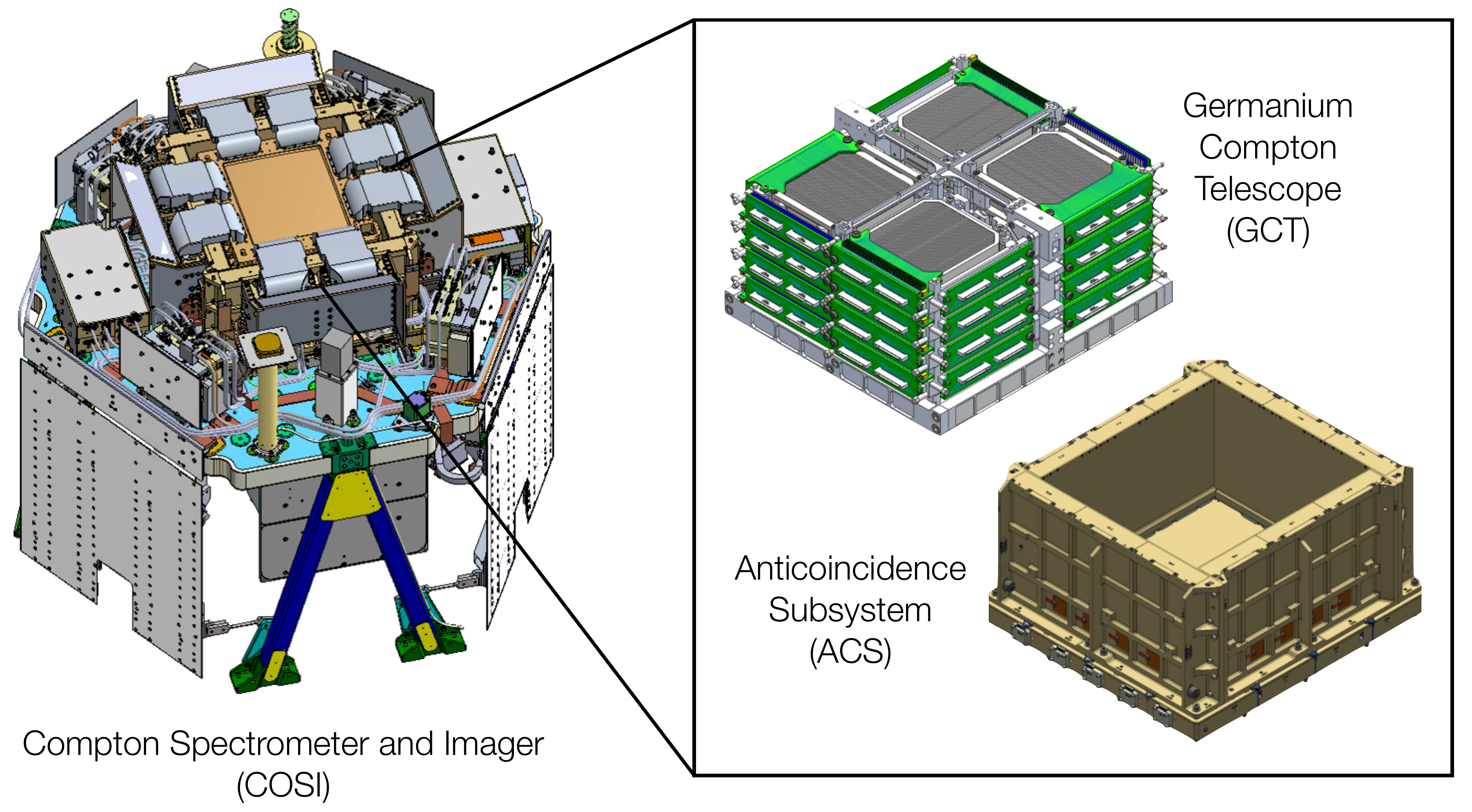}
    \caption{\textit{Left}:~The COSI payload sits on the spacecraft, with the GCT and ACS instruments on the blue payload interface plate deck. \textit{Top right}:~The GCT is a wide field-of-view imaging Compton telescope consisting of 16~cross-strip germanium detectors. \textit{Bottom right}:~The ACS, used for background rejection and transient monitoring, surrounds the GCT on five sides.
    \label{fig:cosi}}
\end{figure*}

The main COSI instrument, the Germanium Compton Telescope (GCT), consists of 16~high-purity germanium detectors shown in Figure \ref{fig:cosi}. The COSI-GCT has an instantaneous field-of-view of $>$25\%~of the sky, enabling wide-field transient monitoring in the 200~keV~$-$~5~MeV regime. The detectors have excellent energy resolution~($\lesssim$1\%) and 3D position resolution~($\sim$1~mm$^3$), making the GCT a sensitive Compton imaging telescope. The GCT event data consist of the simultaneous interactions which occur when an incident gamma ray interacts with the detector, defined by the position and energy. The interactions are reconstructed to determine the energy of each photon and constrain its direction to an annulus on the sky \citep{2022hxga.book...18K}. The GCT can be used for GRB localizations on the order of degrees, spectral analysis, and polarization measurements.

The GCT is surrounded on five sides by a segmented anticoincidence subsystem (ACS) made of bismuth germanium oxide (BGO). The COSI-ACS is sensitive to photon and cosmic ray interactions with energies 80~keV~$-$~40~MeV and instantaneously views the entire Earth-unoccluded sky. It provides active shielding through anticoincidence vetoes, constrains the instantaneous field-of-view of the GCT through passive background reduction, and is used to trigger on transients onboard. Because of its much larger effective area, the ACS will detect many more GRBs than the GCT, and GRBs detected only in the ACS can be localized to~$\sim$$10\degree$ \citep{2026ApJ...997..135P}.

\begin{figure*}[ht!]
    \plotone{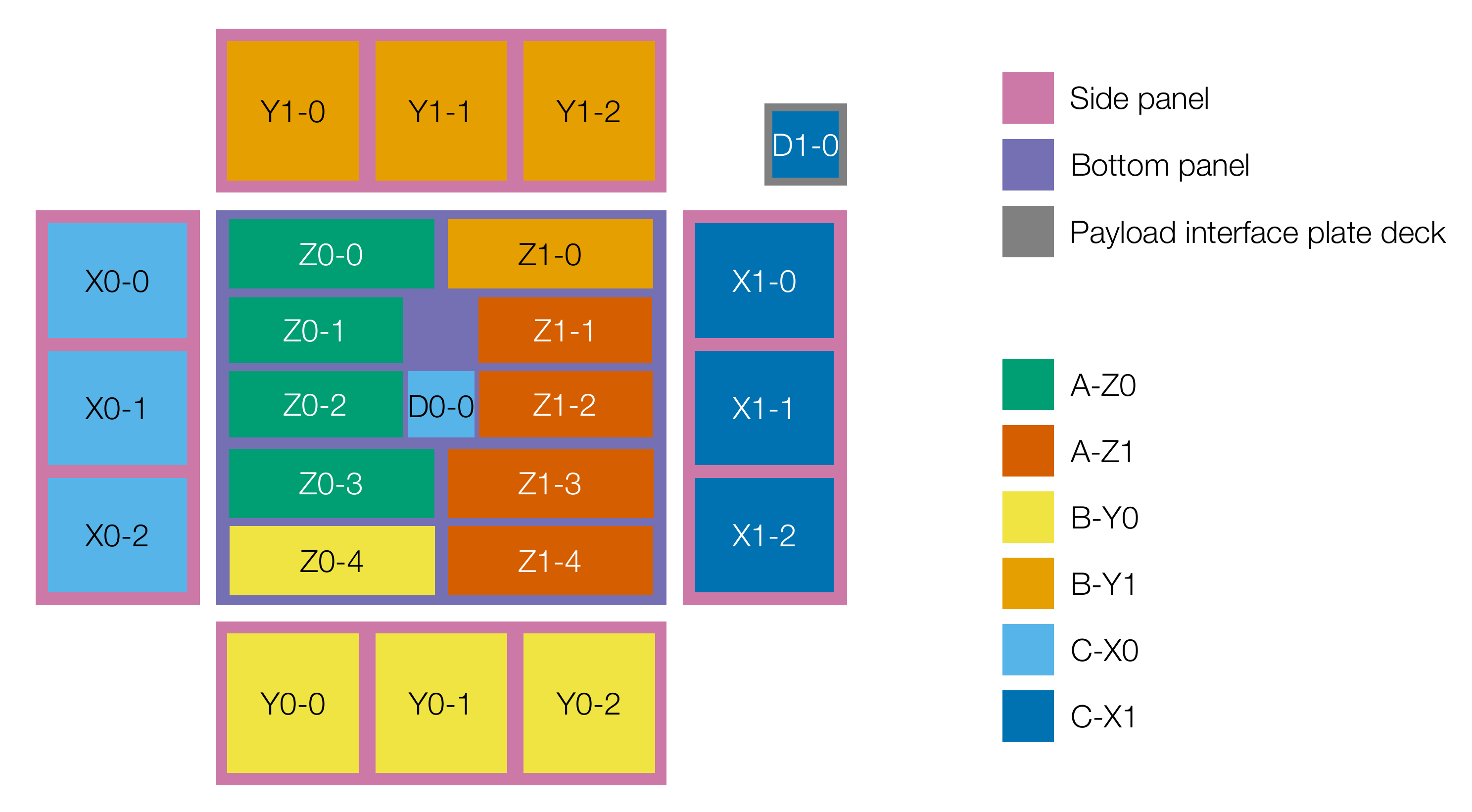}
    \caption{The ACS is comprised of 24~BGO crystals. The ACS binned lightcurve data combines the readout for each of the six~ASICs, color-coded in the figure, with two energy channels (gamma and proton) each, giving 12~channels for the ACS lightcurve data. Note that the B-Y0 and B-Y1 channels combine crystals from the Y-side walls and bottom panel. The D0-0 crystal fills a gap in the clearance hole in the bottom panel that accommodates electrical and thermal connections to the GCT. An identical crystal, D1-0, is placed on the payload interface plate external to the ACS structure.
    \label{fig:acs}}
\end{figure*}

The ACS is made up of 24~BGO crystals, shown in Figure \ref{fig:acs}. Each crystal is read out with an array of silicon photomultipliers (SiPMs), and the signal is split into two channels: gamma (80~keV~$-$~2~MeV) and proton ($2-40$~MeV), where the name is based on the predominant signal contribution. The ACS crystals are designed to surround the GCT with minimal gaps. There is a clearance hole in the bottom panel to accommodate electrical and thermal connections to the GCT array, and a small BGO crystal, D0-0 ($\sim$2.5~cm diameter and $\sim$2.5~cm thickness), is placed to minimize this gap in the bottom panel. There is a second identical BGO crystal, D1-0, on the payload interface plate outside of the ACS for additional background monitoring.

The ACS will transmit two types of data: lightcurve histograms and time-tagged event data. For the lightcurve data, the measured counts from each SiPM are combined into six ACS histogram channels, each with two energy channels (gamma and proton), which largely map to the physical walls of the detector system and are based on the readout Application Specific Integrated Circuits (ASICs). Figure \ref{fig:acs} displays the readout mapping. The lightcurve data will be continuously recorded with 50~ms time bins. A small programmable fraction~($\sim$1\%) of the ACS time-tagged event data will be digitized and telemetered to the ground. The event data consist of 4096~energy channels and identify the crystal number for each interaction.

COSI will typically operate in survey mode, covering the full sky each day by alternately pointing $20\degree$~North of the orbital plane for 12~hours and $20\degree$~South for 12~hours. COSI will launch into a low-Earth equatorial orbit at an altitude of 530~km with $<$$0.4\degree$~inclination. This low-inclination orbit which was chosen to reduce background rates will skim the edge of the South Atlantic Anomaly (SAA), a region of high particle activity. Based on the SAA polygon used for the \textit{Fermi} Gamma-ray Burst Monitor \citep[\textit{Fermi}-GBM;][]{2009ApJ...702..791M}, it is expected that COSI will have increased background rates due to proximity to the SAA such that the onboard trigger algorithm will be disabled for $\lesssim$8\%~of each orbit.

\subsection{Onboard Trigger Algorithm} \label{sec:trigger}

The trigger algorithms onboard COSI are motivated by those used by \textit{Fermi}-GBM \citep{2012ApJS..199...18P,hadar}. \textit{Fermi}-GBM, like many other GRB monitors, runs algorithms onboard the spacecraft to trigger on transients in realtime. This enables the preservation of higher-fidelity data and the automated prompt downlink and processing of data, allowing alerts to be sent to the community and rapid follow-up observations to be performed. The onboard trigger algorithm will be disabled during SAA passages to prevent triggering excessively often on background, between the time between each trigger and the data downlink, and for an adjustable time interval (initially set to 15~minutes) following each trigger to avoid repeatedly triggering on the same transient.

An onboard trigger occurs when the flight software detects a significant increase in measured count rates in the ACS. Onboard COSI, 16~instances of the trigger algorithm, each with its own parameter values, will run concurrently on the measured counts in the 12~ACS binned lightcurve channels. All parameter values are configurable, including the enabled ACS binned lightcurve channels. The instrument triggers a GRB alert if the SNR for any of the 16~realizations of the algorithm exceeds the significance threshold in a configurable number of ACS channels, being $1-12$. We generally require at least two above-threshold channels to prevent triggering on instrumental noise. 

\begin{figure*}[ht!]
    \epsscale{1.15}
    \plottwo{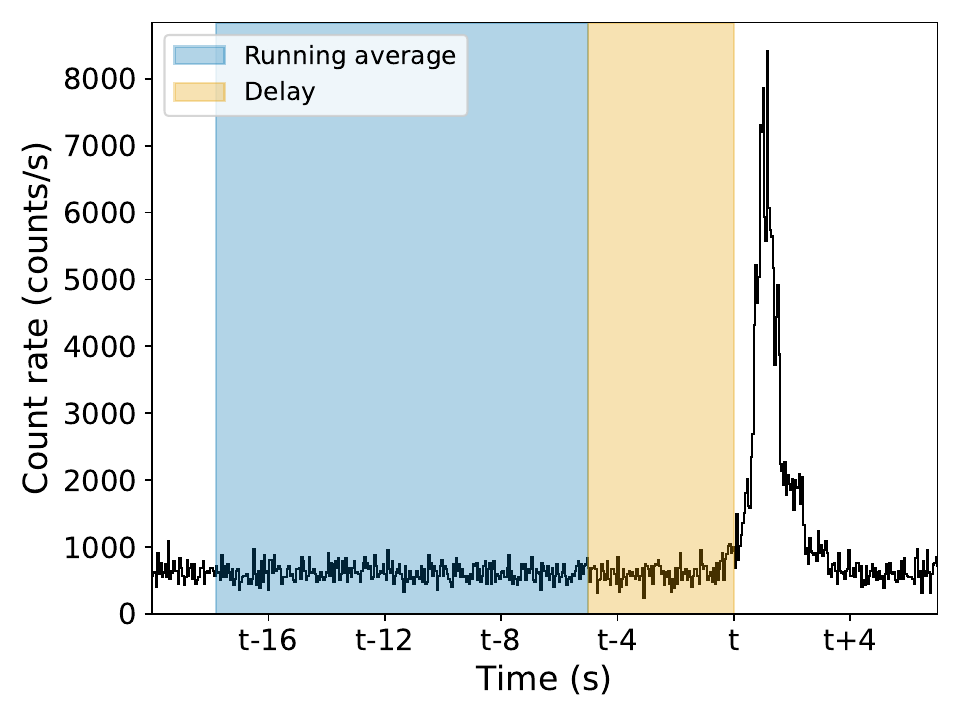}{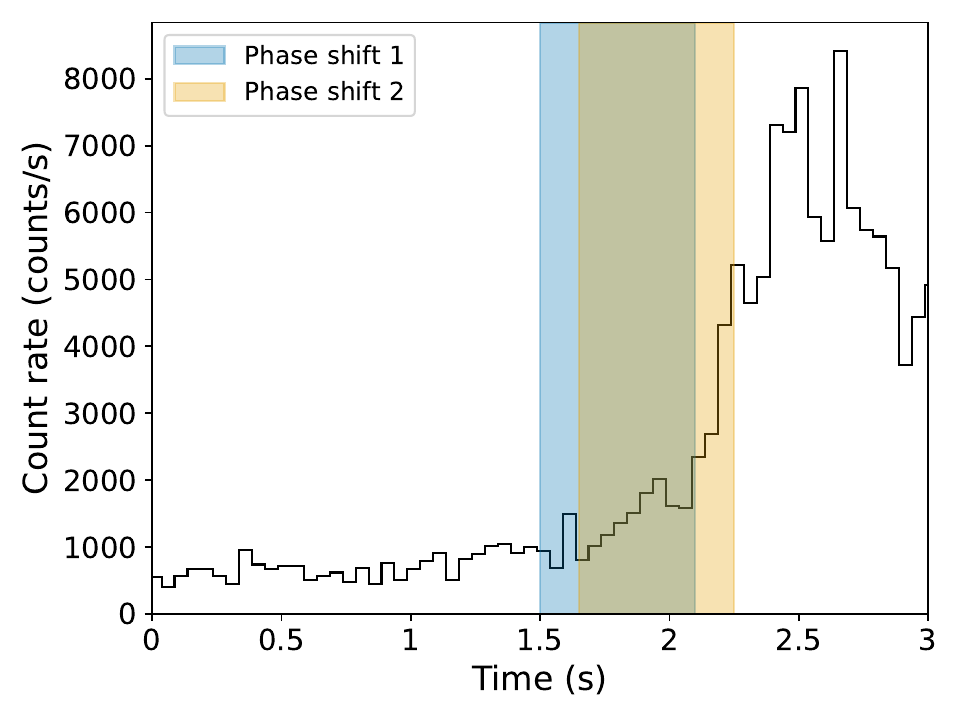}
    \caption{\textit{Left}:~Time interval of 12.8~s defined by $n=8$ used to compute the RA at time $t$ with delay time $D=5$~s. \textit{Right}:~Two instances of trigger algorithm 8 at different phase shifts.
    \label{fig:trigger}}
\end{figure*}

For each of the 12~lightcurve channels, a running average (RA) of the number of counts per 50~ms time bin within a configurable time range $[T_A, T_B]$ is computed as a simple moving average to estimate the background for the SNR calculation. The time range to calculate the RA at time $t$ is given by
\begin{equation}
    [T_A, T_B] = [t - D - (2^n * 50~\text{ms}), t - D].
\end{equation}
where $D$ is a time delay which keeps any potential ongoing burst out of the background estimate and $n$ sets the number of 50~ms time bins used to calculate the RA, both values configurable. For all algorithms in this work, the RA is computed using $D=5$~s and $n=8$, corresponding to 12.8~s. Each algorithm has a configurable phase shift corresponding to the amount of time the search shifts forward between each SNR check. The RA and phase shift are visualized in Figure \ref{fig:trigger}. The total counts ($C$) within a defined and configurable integration time ($T$) are determined, and the SNR is calculated via
\begin{equation}
    \text{SNR} = \frac{C - (\text{RA} * T)}{\sqrt{\text{RA} * T}}.
\end{equation}

The current set of triggers and algorithm parameters, listed in Table \ref{tab:trigger}, are an initial attempt based on simulated data, pursuing low trigger thresholds while minimizing rates of triggers on noise. The algorithm instances were initially outlined using the gamma channel and typically using two detectors, as was done with \textit{Fermi}-GBM. A 50~ms search using three or more detectors was added to target more complete capture of the shortest events. The duration steps also generally favor the shortest times, eventually spanning longer intervals than are active on \textit{Fermi}-GBM, given the greater expected background stability for COSI because of its lower inclination orbit. 

The search was initially run with artificially low thresholds, to capture both a large number of real triggers as well as a robust number of false triggers due to Poisson fluctuations or instrumental effects. The threshold on each algorithm was then increased to exclude all false triggers, by $\sim$0.25$\sigma$ in order to avoid biasing our search by a single set of simulated data. The only false trigger is due to a particularly significant cosmic ray shower which triggered multiple ACS channels on the 50~ms timescale. The trigger algorithm parameter values may be improved ahead of launch. We expect to begin with artificially high thresholds during commissioning, and then iteratively lower the onboard thresholds to empirically determine our acceptable rate of triggers, both from real transients and noise. Currently the six gamma ACS lightcurve channels are enabled for all algorithms, and the proton channels are disabled. The SNR thresholds are currently set to be the same for all lightcurve channels, but they can be configured to be different. 

\begin{deluxetable*}{rllll}
\digitalasset
\tablewidth{0pt}
\tablecaption{COSI trigger algorithm parameters \label{tab:trigger}}
\tablehead{
\colhead{Algorithm} & \colhead{Integration Time (ms)} & \colhead{Channels Over Threshold} & \colhead{Phase Shift (ms)} & \colhead{SNR Threshold}
}
\startdata
0 & 50 & 2 & 50 & 6.6875 \\
1 & 50 & 3 & 50 & 6.1875 \\
2 & 100 & 2 & 50 & 6.5625 \\
3 & 150 & 2 & 50 & 6.4375 \\
4 & 200 & 2 & 50 & 6.3125 \\
5 & 250 & 2 & 50 & 6.1875 \\
6 & 300 & 2 & 100 & 6 \\
7 & 400 & 2 & 100 & 5.875 \\
8 & 600 & 2 & 150 & 5.875 \\
9 & 800 & 2 & 200 & 5.875 \\
10 & 1600 & 2 & 400 & 6 \\
11 & 2400 & 2 & 600 & 6.1875 \\
12 & 3200 & 2 & 800 & 6.875 \\
13 & 4800 & 2 & 1200 & 6.875 \\
14 & 6400 & 2 & 1600 & 7.6875 \\
15 & 12800 & 2 & 3200 & 5.75 \\
\enddata
\tablecomments{The COSI onboard trigger algorithm values used in this work. These parameters are configurable up to and after launch. Currently the six gamma ACS lightcurve channels are enabled for all algorithms, and the proton channels are disabled. The SNR thresholds are currently the same for all ACS lightcurve channels, but this is configurable on a per-channel basis. For all algorithms, the RA is computed using $D=5$~s and $n=8$, corresponding to an RA integrated over 12.8~s.}
\end{deluxetable*}

Following an onboard trigger, a limited dataset will be downlinked within an hour during brief scheduled downlinks or space relay contacts. The received 500~kbit GRB alert data packets will contain rebinned ACS lightcurve data for a minute-long interval around the trigger time $T_0$, information about the triggered algorithm, housekeeping data including live times and spacecraft position and orientation, and as many GCT Compton events as can fit in the data packets ($\sim$600). The ACS lightcurve data contained in the GRB alert data will include both the gamma and proton energy channels for one minute surrounding the trigger with the time binning listed in Table~\ref{tab:acs-binning}. The duration of some long~GRBs will exceed the duration of the ACS data downlinked in the GRB alert data, meaning that the full event will not be captured in the GRB alert data and there will be no background-only lightcurve bins following the burst. High-latency data surrounding the trigger, including GCT event data, pre-scaled ACS event data, and the full 50~ms ACS binned lightcurves, will be saved onboard and downlinked during high data-rate ground passes every $\sim$3~hours. We aim to have scheduled low data-rate ground passes every $\sim$$30-60$~minutes, enabling the downlink of the GRB alert packets within an hour of the burst. We assume this longer 60~minute cadence in our reported sGRB rates to be conservative.

\begin{deluxetable*}{rllll}
\digitalasset
\tablewidth{0pt}
\tablecaption{ACS data binning of GRB alert packet \label{tab:acs-binning}}
\tablehead{
\colhead{Time Interval Start (s)} & \colhead{Time Interval End (s)} & \colhead{Time Bin Size (ms)}
}
\startdata
$T_0-30$ & $T_0-10$ & 1000 \\
$T_0-10$ & $T_0-1$ & 250 \\
$T_0-1$ & $T_0+4$ & 50 \\
$T_0+4$ & $T_0+30$ & 250 \\
\enddata
\tablecomments{Time binning of both the gamma and proton ACS data channels in the GRB alert packet downlinked following an onboard trigger, where $T_0$ is the trigger time.}
\end{deluxetable*}

\subsection{Transient Analysis Pipelines} \label{sec:transient-pipelines}

Analysis of transients observed by COSI will be performed using two main pipelines running at the Science Data Operations Center: the Fast Transient Analysis Pipeline (FasTP) and the Comprehensive Transient Pipeline. The FasTP processes the GRB alert packet data from onboard triggers, whereas the Comprehensive Transient Pipeline analyzes the data telemetered during high data-rate ground passes $\sim$7~times per day. Both pipelines are implemented using the COSIflow framework \citep{10.1117/12.3105168} built on Apache Airflow and use the COSI high-level analysis software \texttt{cosipy} \citep{2024icrc.confE.858M,the_compton_spectrometer_and_imager_cosi_2025_15125776}. 

The FasTP will localize \citep{2026ApJ...997..135P,2026arXiv260415119P}, classify, and perform spectral analysis of GRBs using the GCT and ACS data in a fully automated way. Results determined by the FasTP will be disseminated to the scientific community within an hour of the burst detection via the General Coordinates Network \citep[GCN; ][]{2000AIPC..526..731B} and NASA's High Energy Astrophysics Science Archive Research Center (HEASARC). The scientific analysis tools implemented for the FasTP are tailored to the GRB alert data packet with the binning schema described in Section \ref{sec:trigger}.

The Comprehensive Transient Pipeline will perform localizations, spectral analysis, and polarization measurements of GRBs using the high-latency data saved during a trigger, as well as conduct subthreshold searches for transients. It can improve the results obtained by the FasTP, which is limited by low data-rate telemetry data file size constraints. After preliminary results are disseminated to the community based on FasTP analysis of the GRB alert data packet, a comprehensive human-in-the-loop GCN circular will be distributed with more detailed information about the GRB determined by the Comprehensive Transient Pipeline.

\section{Simulated COSI sGRB Observations} \label{sec:simulations}

We simulate COSI observations of sGRBs in order to determine the detection efficiency. Using 170~spectral models and lightcurves based on \textit{Fermi}-GBM observations of sGRBs, we run simulations of 2720 bursts positioned throughout the COSI field-of-view. These simulated COSI sGRB observations are then run through the onboard trigger algorithm, and we analyze the resulting trigger efficiency. In this paper, we describe the performance of the trigger algorithm and the sGRB detection rate with the ACS, while future publications will detail the burst localization capabilities and science validation of the FasTP and Comprehensive Transient Pipeline.

\begin{figure*}[ht!]
    \epsscale{1.15}
    \plottwo{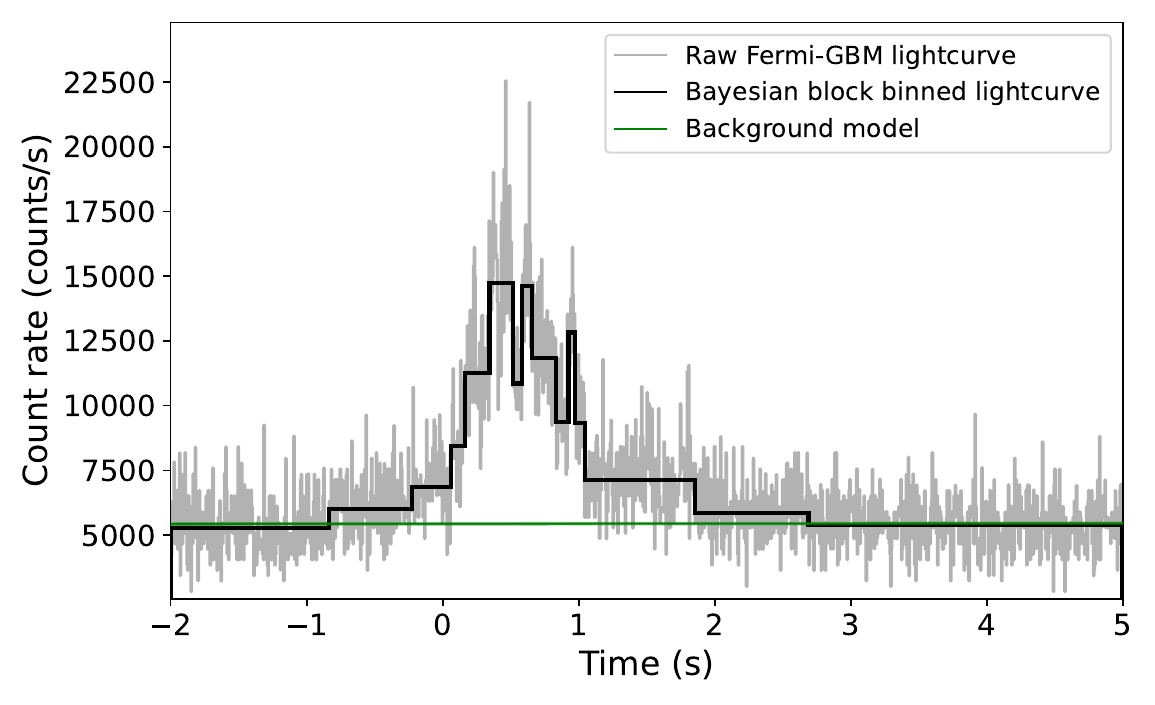}{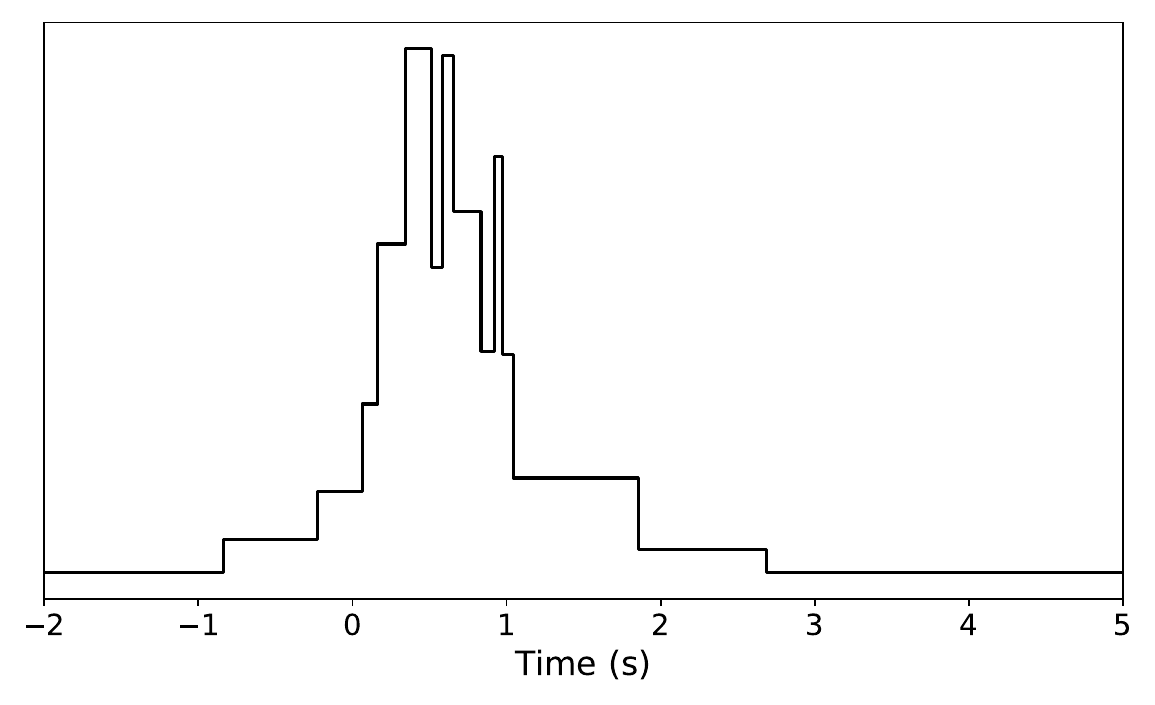}
    \caption{\textit{Left}:~Example Bayesian block binning of the \textit{Fermi}-GBM observation of GRB~240216A, including the fitted background. \textit{Right}:~Background-subtracted lightcurve of GRB~240216A used as input for COSI simulations. The amplitude is arbitrary and set by the peak flux chosen for the simulation.
    \label{fig:sgrb-input-lightcurve}}
\end{figure*}

To elaborate, we create a library of sGRB models using \textit{Fermi}-GBM data to simulate with COSI. As input models, we use lightcurves and time-integrated spectra of GRBs with $T_{90}<2.5$~s observed by \textit{Fermi}-GBM between July~23,~2008 and September~13,~2024 \citep{2020ApJ...893...46V}. While sGRBs are typically defined by having $T_{90}<2$~s, this does not encompass all GRBs from binary neutron star mergers, including GRB~170817A with \textit{Fermi}-GBM $T_{90}=2.048$~s. Based on the Fourth \textit{Fermi}-GBM GRB catalog \citep{2020ApJ...893...46V}, 80.6\% of GRBs with $T_{90}=2.5$~s belong to the sGRB population if the $\log{T_{90}}$ distribution is modeled as two Gaussian distributions. We limit our sample of \textit{Fermi}-GBM GRBs used to produce our input models to have a minimum $50-300$~keV peak flux of 5~ph/cm$^2$/s and a time-integrated spectrum best fit with a Band or Comptonized function, leaving 170~sGRB models. This is done to only include bursts with enough statistics to capture the temporal structure and a realistic spectrum with constrained curvature. Using the Fermi Gamma-ray Data Tools \citep{GDT-Fermi}, we bin the \textit{Fermi}-GBM lightcurve in the 10~keV~$-$~10~MeV energy band using Bayesian blocks with false alarm probability $p_0=0.005$. We estimate the background for each GRB by fitting a polynomial to the lightcurve in the $[T_0-22, T_1-1]$~s and $[T_2+1, T_0+25]$~s time ranges where $T_0$ is the trigger time and $[T_1, T_2]$ is the $T_{90}$ interval. The time intervals were chosen through trial and error to produce successful background fits for all bursts through an automated fitting procedure. Then, we subtract the fitted background to produce a lightcurve for each GRB. We are left with 170~unique sGRB models consisting of a lightcurve and spectrum. Figure \ref{fig:sgrb-input-lightcurve} shows an example of a \textit{Fermi}-GBM lightcurve binned using Bayesian blocks used as input for the COSI simulations.

\begin{figure*}[ht!]
    \epsscale{1.15}
    \plottwo{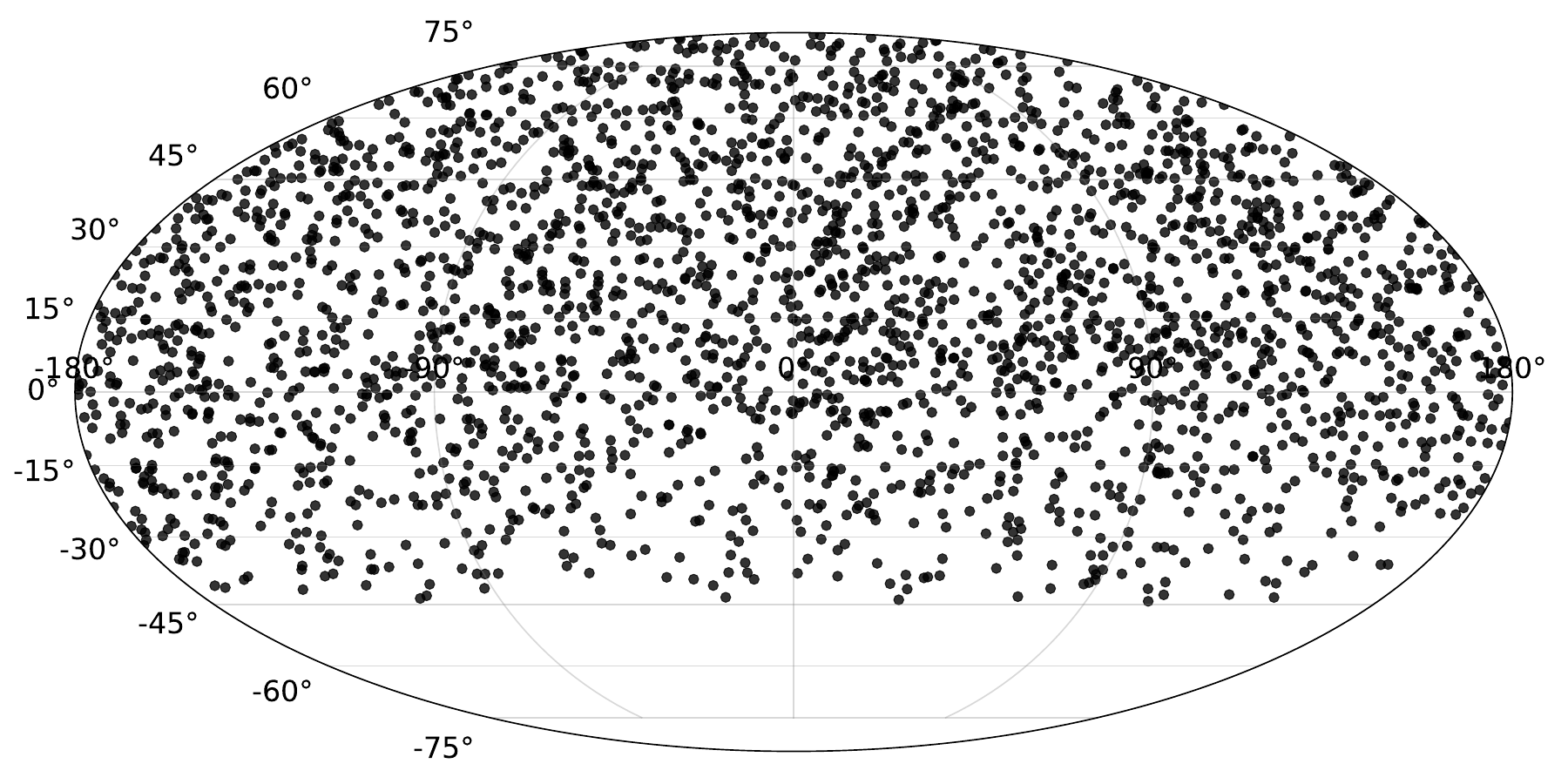}{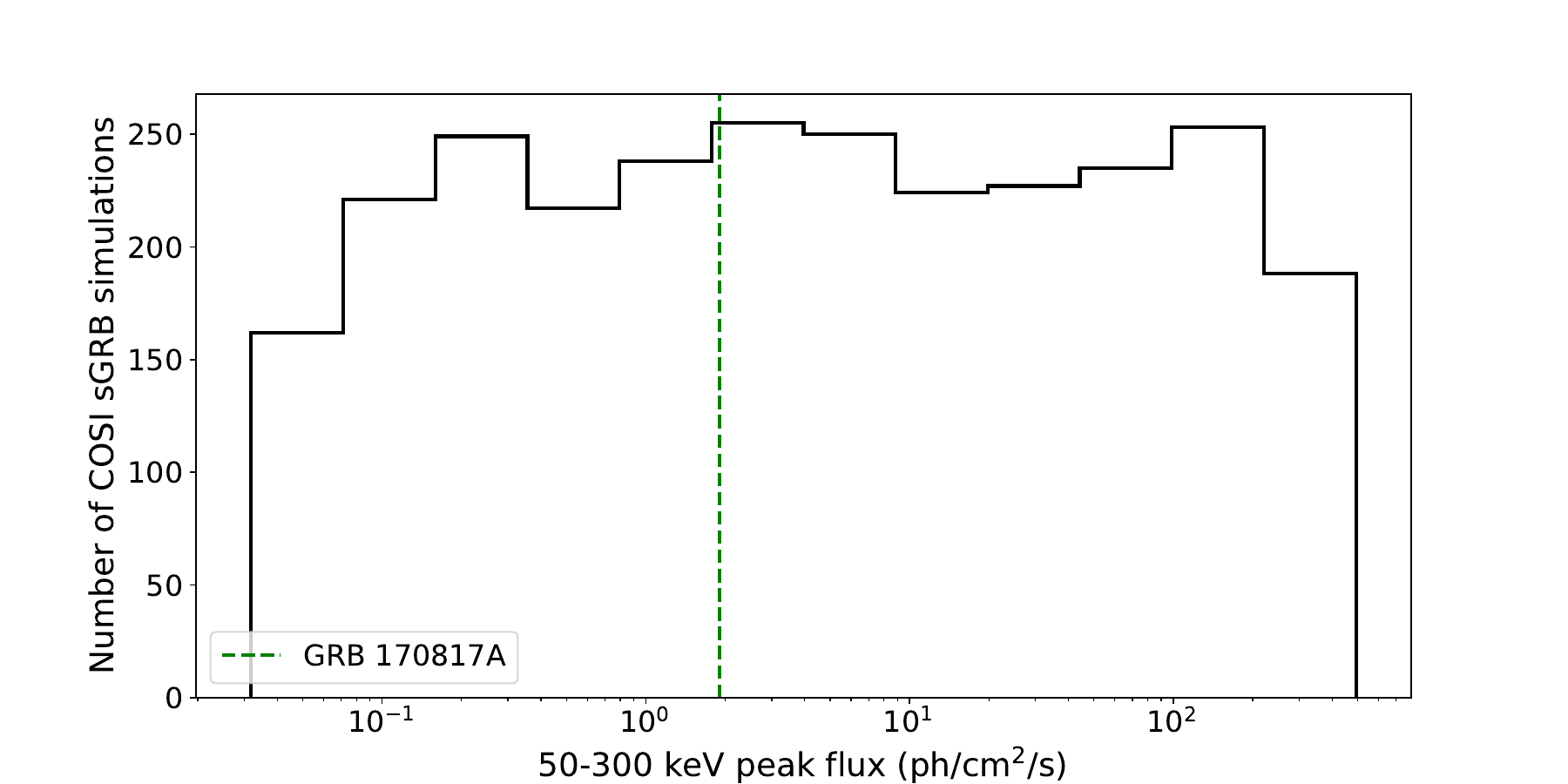}
    \caption{\textit{Left}:~Position distribution of simulated COSI sGRBs in spacecraft coordinates. \textit{Right}:~$50-300$~keV peak flux distribution for COSI sGRB simulations, with the 64~ms peak flux of GRB~170817A shown for comparison.
    \label{fig:cosi-input-distributions}}
\end{figure*}

We simulate 2720~COSI observations of these sGRB models between 10~keV~and~10~MeV using \texttt{cosiburstpy} \citep{cosiburstpy}, which produces population-level COSI simulations of transient observations. The GRBs are uniformly distributed on the sky excluding the region blocked by the Earth and times when COSI is within the SAA or slewing. Each of the 170~models is simulated 16~times with the 10~keV~$-$~10~MeV peak fluxes log-uniformly distributed between 0.1~and~1000~ph/cm$^2$/s. There is a fixed pairing of the lightcurve and spectrum for each GRB input model according to the \textit{Fermi}-GBM data (i.e. we do not mix and match). Figure \ref{fig:cosi-input-distributions} shows the position and $50-300$~keV peak flux distributions. The simulations are performed using the Medium Energy Gamma-ray Astronomy Library (MEGAlib), a GEANT4 wrapper developed to simulate and analyze MeV gamma-ray instrument data \citep{2006NewAR..50..629Z}. MEGAlib simulates the interactions and of incident photons from an astrophysical source using a detailed mass model of COSI, and outputs the energies and positions of the interactions in the GCT and ACS. The spacecraft orbit and orientation, including Earth occlusion, are modeled in MEGAlib as described in \citet{2026ApJ...997..284G}. For these simulations, COSI rocks between $22\degree$~North for 12~hours and $22\degree$~South for 12~hours with an 8~minute slewing time as described in \citet{karwin_2026_20602657}.

\begin{figure*}[ht!]
    \plotone{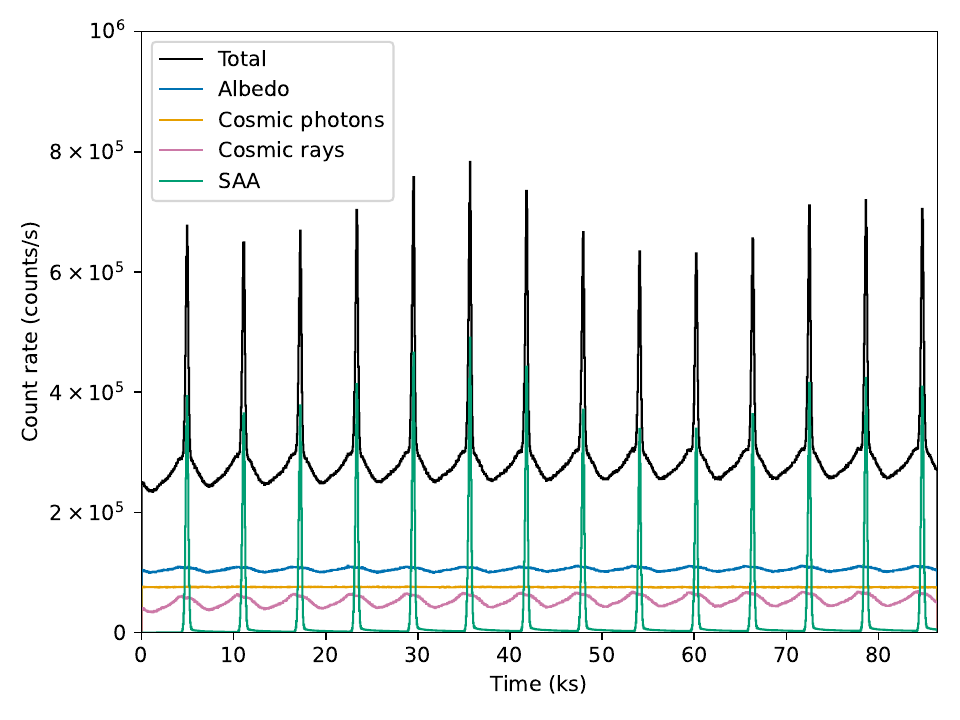}
    \caption{COSI-ACS background in the gamma channel for all ASICs combined during the first 24~hours after launch. The simulations are described in \citet{2026ApJ...997..284G}, with an 18\%~increase in flux used for this work to represent the expected background after two years in orbit.
    \label{fig:background}}
\end{figure*}

\begin{figure*}[ht!]
    \plotone{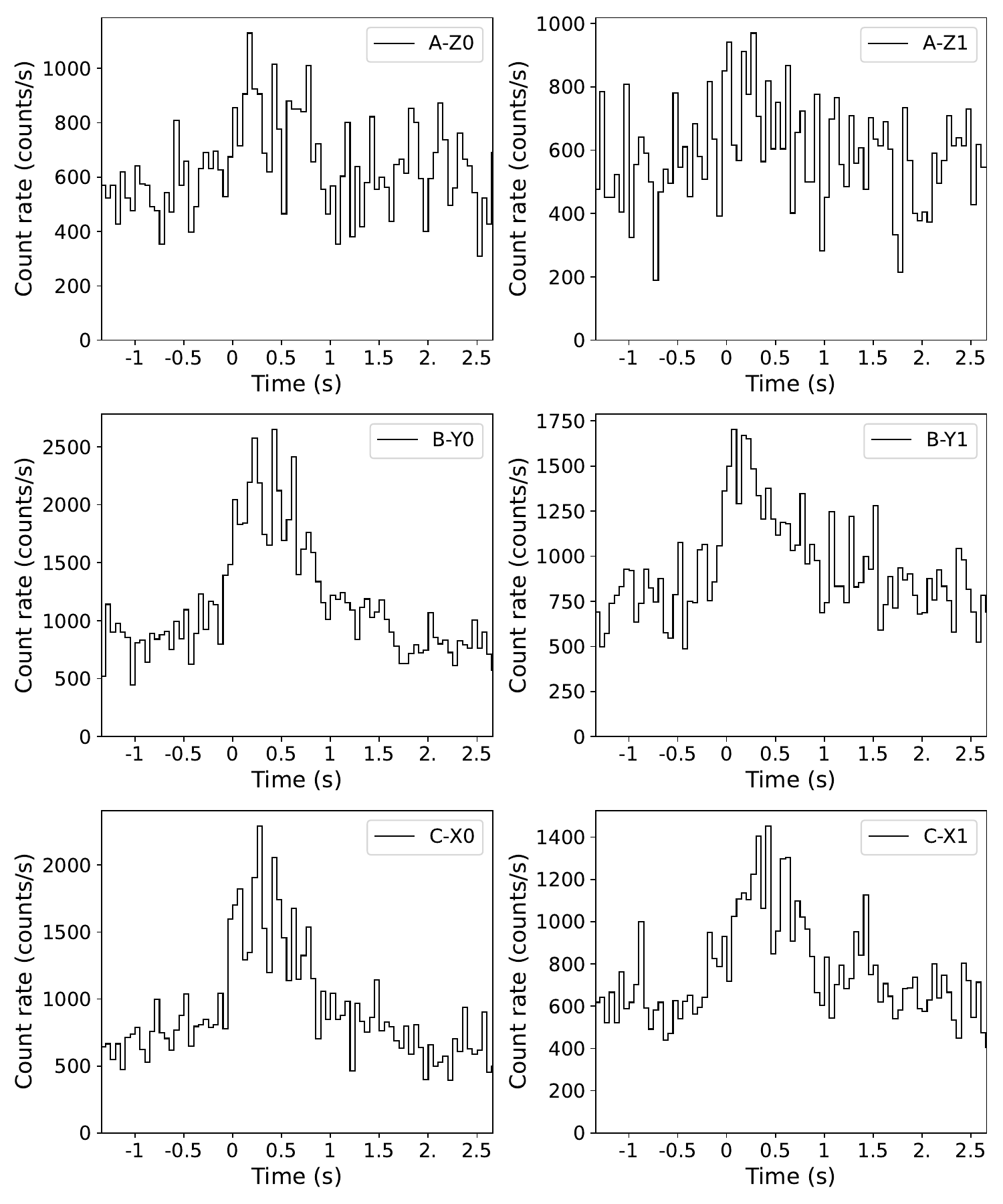}
    \caption{Example COSI-ACS gamma channel lightcurves of a GRB observation including background for each of the 6~ASICs with 50~ms resolution. The input model for the simulated COSI observation is the lightcurve (shown on the right side of Figure \ref{fig:sgrb-input-lightcurve}) and spectrum of GRB~240216A with a $50-300$~keV peak flux of~2.8~ph/cm$^2$/s, incidence angle of~$68.7\degree$, and azimuthal angle of~$323.96\degree$.
    \label{fig:acs-sgrb-lightcurve}}
\end{figure*}

We combine the simulated COSI sGRB observations with ACS background, shown in Figure \ref{fig:background}, during the first 24~hours after launch based on the background simulations reported in \citet{2026ApJ...997..284G}. The background includes Galactic and extragalactic background photons, albedo glow from Earth’s atmosphere, cosmic rays, the SAA, and the resulting decay of unstable isotopes in the instrument. The background rate due to the decay of isotopes increases over the mission lifetime as unstable isotopes build up in the instrument. In the background simulations from \citet{2026ApJ...997..284G}, the 100~keV~$-$~2~MeV Compton event rate increases by $\sim$18\% between the first day and two years in orbit. Because the ACS background simulation used here is for the first 24~hours after launch, we increase the counts by 18\% to represent the expected background after two years in orbit. This artificially increases the Poisson fluctuations of the measured counts by $\sim$10\% ($\sqrt{1.18}=1.09$), which we retain to be conservative. Figure \ref{fig:acs-sgrb-lightcurve} shows an example of the measured ACS lightcurve in the six ACS gamma channels of the same GRB as in Figure \ref{fig:sgrb-input-lightcurve}, including the simulated background. 

In order to estimate the rate of sGRBs which can be well-localized using the GCT, the sGRB Compton events are reconstructed from the raw interactions and combined with the background discussed in \citet{2026ApJ...997..284G}, with an 18\%~increase in normalization as with the ACS data. Event selections are performed as discussed in \citet{karwin_2026_20602657}. 

We note that the small D0-0 and D1-0 crystals~($\sim$5~cm$^2$ active area) shown in Figure \ref{fig:acs} are not included in the simulations presented here; however, their impact is minimal to the trigger algorithm results due to their small size compared to the total active area of the A-X0 and A-X1 readout channels~(687 cm$^2$ each). 

\section{sGRB Detection Rate and Gravitational Wave Coincidences} \label{sec:sgrb-rate}

\begin{figure*}[ht!]
    \plotone{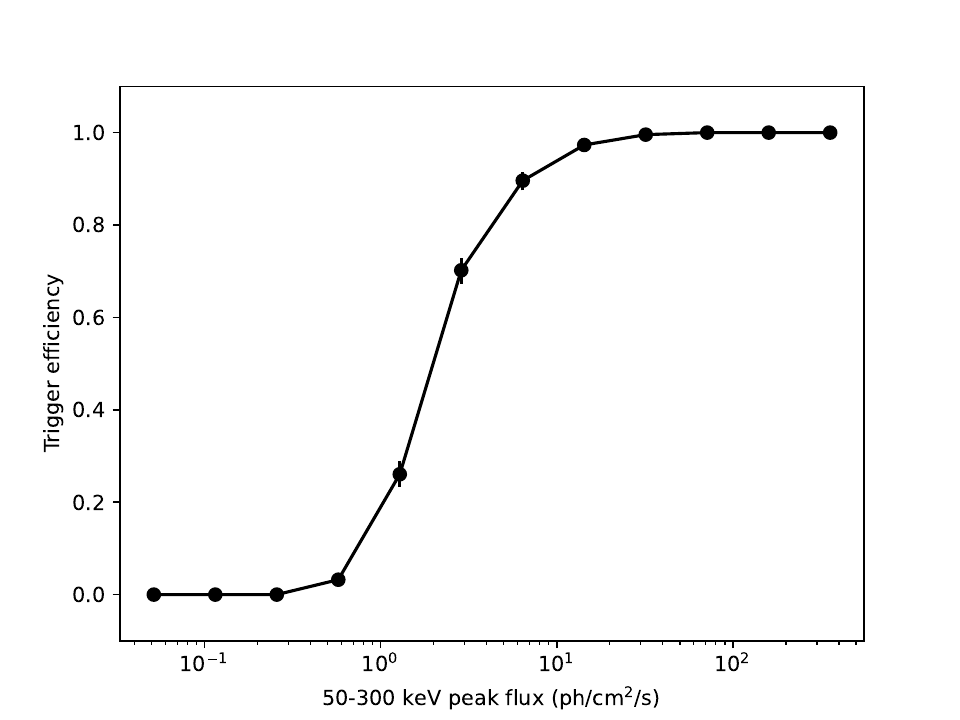}
    \caption{COSI sGRB trigger efficiency as a function of peak flux. COSI achieves a $>$50\% trigger efficiency for $50-300$~keV peak fluxes~$\gtrsim$1.5~ph/cm$^2$/s, $>$75\% trigger efficiency for peak fluxes~$\gtrsim$2.5~ph/cm$^2$/s, and 100\% trigger efficiency for peak fluxes~$\gtrsim$30~ph/cm$^2$/s. The error bars are the~1$\sigma$ binomial uncertainty.
    \label{fig:efficiency-flux}}
\end{figure*}

\begin{figure*}[ht!]
    \epsscale{1.15}
    \plottwo{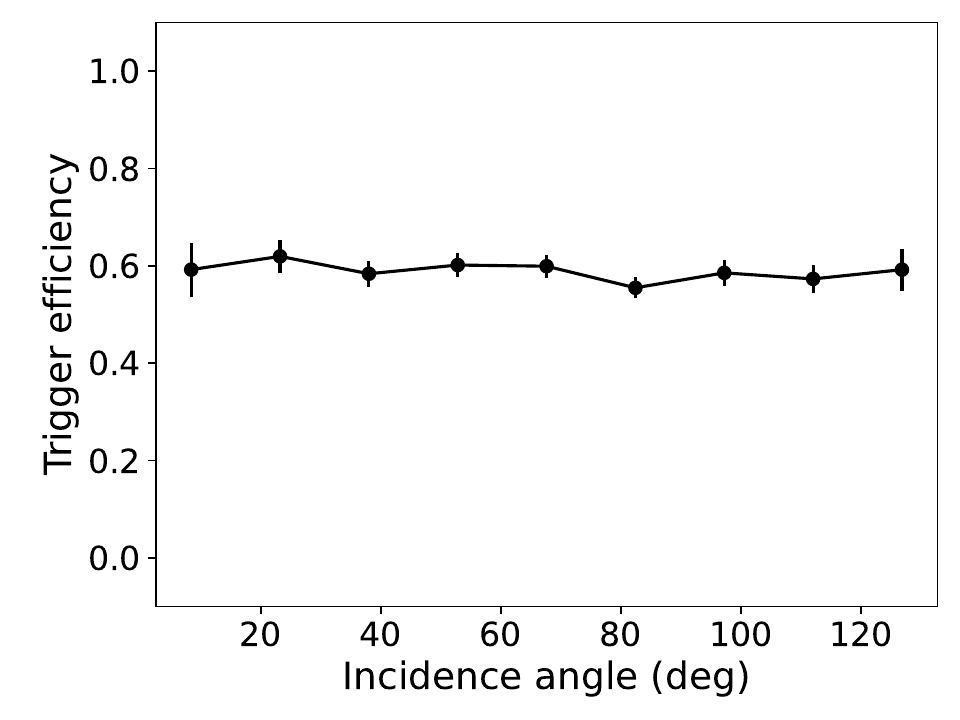}{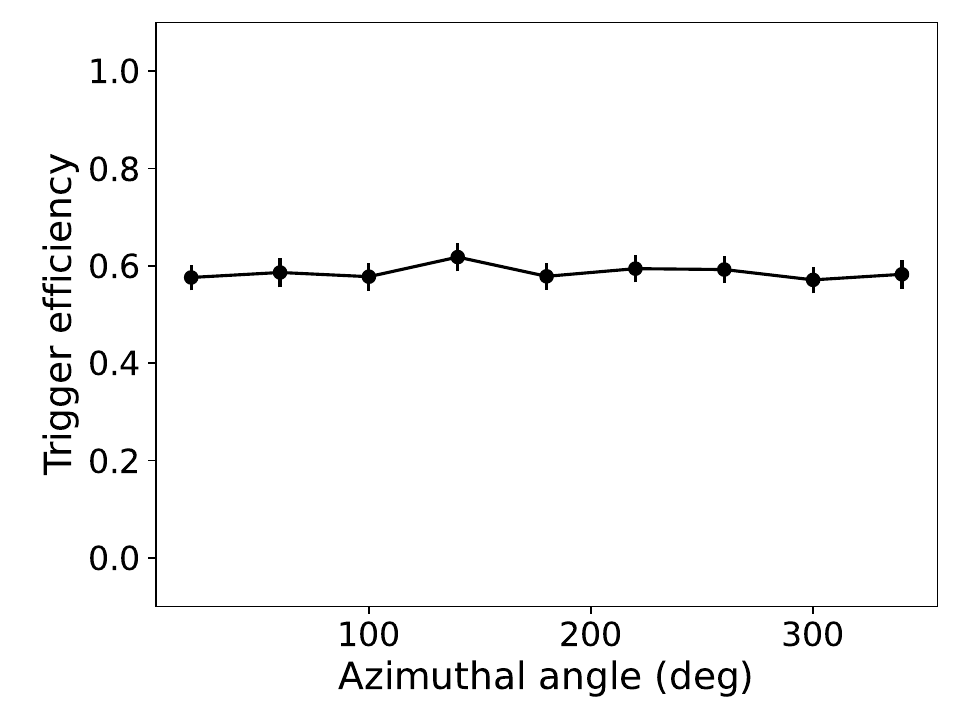}
    \caption{The COSI-ACS has approximately uniform sensitivity to GRBs across the field-of-view for 10~keV~$-$~10~MeV peak fluxes between $0.1$~and~$1000$~ph/cm$^2$/s. \textit{Left}:~COSI trigger efficiency as a function of sGRB incidence angle. \textit{Right}:~COSI trigger efficiency as a function of sGRB azimuthal angle.
    \label{fig:efficiency-position}}
\end{figure*}

We pass the simulated ACS data through a Python emulation of the onboard trigger algorithm to estimate the sGRB trigger efficiency, or fraction of simulated sGRBs that triggered the instrument. Figure \ref{fig:efficiency-flux} shows the trigger efficiency as a function of $50-300$~keV peak flux. The COSI-ACS provides uniform sensitivity to bursts throughout the entire Earth-unoccluded sky, as displayed in Figure \ref{fig:efficiency-position}.

\begin{figure*}[ht!]
    \plotone{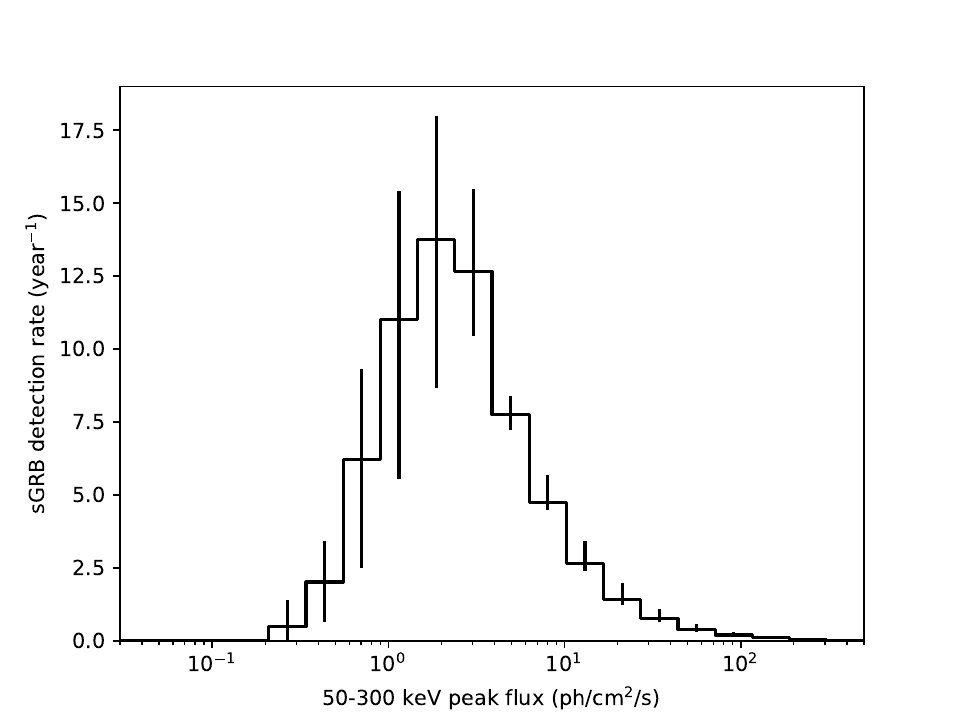}
    \caption{COSI sGRB detection rate as a function of peak flux. Integrating over all peak flux bins, COSI is expected to detect 68~sGRBs per year, with a 90\% confidence range spanning~$47-89$.
    \label{fig:sgrb-detection-rate}}
\end{figure*}

To predict the rate of COSI sGRB detections, we convolve the intrinsic sGRB rate in the universe described in E. Burns (2026, in preparation) with the COSI onboard trigger efficiency. In brief, the technique fits the cumulative distribution of \textit{Fermi}-GBM peak fluxes on the 64~ms timescale for which \textit{Fermi}-GBM has near 100\% detection efficiency against a large set of simulated sGRB populations arising from various combinations of star formation rate, neutron star merger evolution, intrinsic luminosities, jet structures, etc. The outcome is a range of all-sky sGRB rates as a function of 64~ms peak flux. This is convolved with the calculated COSI trigger efficiency curve, sampling the uncertainty in each peak flux bin, to determine the COSI detection rate. Because the COSI trigger efficiency curve shown in Figure \ref{fig:efficiency-flux} assumes that the sGRBs are within the ACS field-of-view and occur outside of the SAA, we scale the COSI detection rates assuming 30\%~sky occlusion by Earth at an orbital height of 530~km, 8\%~downtime during SAA passages, and losses due to the disabling of the onboard trigger algorithm between the time of the trigger and data downlink, which we treat as a uniform distribution between two minutes and one hour. We achieve an expected detection rate of 68~sGRBs per year, with a 90\%~confidence range spanning~$47-89$. The predicted detection rate as a function of peak flux is shown in Figure \ref{fig:sgrb-detection-rate}. For comparison, \textit{Fermi}-GBM observes $\sim$$45-55$~sGRBs per year \citep{2016ApJ...818..110B,2020ApJ...893...46V}. 

With this method, we use one simulated GRB population to determine the detection efficiency of COSI, and a separate population starting from intrinsic properties to determine observed sGRB rates when convolved with the COSI trigger efficiency curve. These two underlying populations would ideally be a single dataset which would account for correlations between properties, such as the spectral hardness and flux. However, we set our simulated population used for detection efficiency to use observed lightcurves in order to best inform the set of trigger algorithms running onboard COSI, and there is no clear method to sample lightcurves with a forward-modeled GRB population. As such, ours is decoupled. After comparing the properties of the \textit{Fermi}-GBM GRBs which we use as input models for our COSI simulations against the population of GRBs using intrinsic properties and ranges of rate evolution, we find no strong evidence of a biased sample and expect $\lesssim$10\%~variations.

As a sanity check on our predicted detection rate, we scale the rate of \textit{Fermi}-GBM sGRB detections to estimate the ACS detection rate. We calculate all-sky sGRB rates from the effective \textit{Fermi}-GBM trigger threshold and detection rate, accounting for the observing fraction of~0.596 because of losses due to Earth-occlusion, downtime due to SAA passages, and deadtime following onboard triggers. This is scaled by the relative trigger threshold values between \textit{Fermi}-GBM and COSI, relative sizes of the crystals~($\sim$3900~cm$^2$ total for the ACS and $\sim$127~cm$^2$ for each of the 12~\textit{Fermi}-GBM sodium iodide crystals), and effective area over the sky for the second-best viewing detector. This gives a fiducial rate of 45~sGRBs per year for COSI; however, this does not account for the lower predicted background of COSI due to its decreased time spent in the SAA and therefore is a lower limit.

Using the same 2720~simulated COSI sGRB observations, we explore the rate of sGRBs that can be well-localized using the GCT. Specifically, we use \texttt{cosipy} to calculate the SNR for each simulated sGRB and use the assumption that sGRBs with SNR~$>6$ are expected to be localized to within~$2.5\degree$, based on the COSI requirement. The SNR is computed via
\begin{equation}
    \text{SNR} = \frac{S}{\sqrt{S+B}}
\end{equation}
where $S$ are the simulated GCT source counts for a given GRB and $B$ are the simulated background counts during the duration of the GRB. We then convolve the fraction of simulated sGRBs detected in the ACS with SNR~$>6$ in the GCT as a function of peak photon flux with the ACS detection rate. Based on this, 8~sGRBs per year with a 90\%~confidence range spanning $\sim$$7-10$ are expected to have SNR~$>6$ in the GCT and therefore based on the COSI requirement, be localized to~$<$2.5$\degree$. This is a much lower number than will be detected in the ACS due to the smaller effective area and field-of-view, efficiency of Compton reconstruction, and more limited energy range. However in addition to superior localizations, the GCT data can be used for high-resolution spectral studies and polarization analysis.

The formalization for sGRB rates as a function of brightness captures redshift-dependent rate evolution, which allows an estimate of the joint GW-GRB detection rate. We take the expected GW sensitivity curve for the O5c run and the nominal design sensitivity for the A+ detectors from \citet{Fritschel2025LIGO} to generate GW detection efficiency as a function of distance for face-on, canonical 1.4+1.4~M$_\odot$~binary neutron star mergers using the inspiral-range package \citep{chen2021distance}. We assume two LIGO interferometers, each with 70\%~livetime. We require a network SNR of 9.6 when both interferometers are active and a single interferometer SNR of~12 when only one is active, relying on the spatial and temporal alignment with the GRB for robust confirmation of these events. The GW sensitivity curve is then convolved with the absolute rate of COSI detections as a function of distance over the range of viable rates models. This gives a resulting GW-GRB rate with COSI at our simulated sensitivities of 0.8 events per year, with a 90\%~confidence range spanning~$0.4-2.3$ which, when including Poisson rate uncertainty, corresponds to $0-3$~joint detections per year.

\section{Conclusion} \label{sec:conclusion}

COSI is an upcoming gamma-ray space telescope with a two-year nominal mission beginning in 2027. The COSI-ACS will trigger on transients onboard the spacecraft and downlink GRB alert data within $\sim$$30-60$~minutes. An alert including a localization will be disseminated to the community within an hour of the burst via GCN in order to enable prompt multi-wavelength follow-up of the afterglow.

This paper presents a description of the onboard trigger algorithm and quantifies the ACS trigger efficiency. We simulate 2720 COSI observations of sGRBs using input models derived from \textit{Fermi}-GBM sGRB observations. These simulations are passed through an emulation of the onboard trigger algorithm, which searches for significant count rate increases in at least two of the 12~ACS binned lightcurve channels, to determine the onboard trigger efficiency of COSI as a function of peak photon flux. 

This COSI onboard trigger efficiency is convolved with an intrinsic sGRB rate in the universe to estimate that COSI will detect $\sim$$47-89$~sGRBs per year. A short-to-long GRB ratio of~$\sim$30\% is expected for COSI (E. Burns 2026, in preparation), so we predict $\sim$157~$-$~297~long GRB detections per year and $\sim$204~$-$~386~total onboard triggers annually on GRBs. 

The COSI-GCT will be used to localize GRBs and perform spectroscopy and polarimetry. To estimate the rate of well-localized sGRBs, we calculate the SNR of the sGRB simulations in the GCT and convolve the fraction of detected GRB simulations with SNR~$>$~6 with the calculated ACS detection rate. We predict that the brightest $\sim$$7-10$~sGRBs per year will be localized to within~2.5$\degree$.  

The O5 LIGO-Virgo-KAGRA interferometer run is scheduled to begin in 2029, during which time it will be possible to jointly detect GWs and sGRBs detected by COSI during its probable extended mission phase. Using the predicted O5 GW detection efficiency as a function of distance for binary neutron star mergers and the expected ACS detection rate, we find that the anticipated GW-GRB rate with COSI is $0-3$~per year. Even a single joint observation between a COSI GRB and GW event would double the population of coincident detections and could be used to probe sGRB progenitors, jet structure and composition, and emission mechanisms.

\begin{acknowledgments}

We thank M. Moss for insightful discussions. 

The Compton Spectrometer and Imager is a NASA Explorer project led by the University of California, Berkeley with funding from NASA under contract 80GSFC21C0059. Resources supporting this work were provided by the NASA High-End Computing (HEC) Program through the NASA Advanced Supercomputing (NAS) Division at Ames Research Center. EN acknowledges NASA funding through Cooperative Agreement 80NSSC22K0982. This work is supported by the Italian Space Agency, partially within contract ASI/INAF No. 2024-11-HH.0. The material is based upon work supported by NASA under award number 80GSFC24M0006.

\end{acknowledgments}

\begin{contribution}

E. Neights led the analysis and writing of the paper, analyzed the \textit{Fermi}-GBM data, ran and processed the COSI sGRB simulations, and calculated the rate of well-localized sGRBs. E. Burns contributed to the writing, wrote an emulation of the trigger algorithm and ran it on the COSI simulations, and calculated the sGRB and GW rates. N. Parmiggiani contributed to the wiring and discussions of the results. Y. Sheng analyzed the GCT data. S. Gallego simulated the COSI background. A. Aguilar-Nadalini wrote the flight software for the trigger algorithm. A. C. Trigg helped analyze the \textit{Fermi}-GBM data. S. Guiriec helped refine the text and contributed scientific expertise. Authors K. Abhijith Augustine through P. Veres are listed alphabetically and contributed relevant scientific expertise. C. A. Kierans assisted with interpreting the results and refining the text. Authors S. E. Boggs through A. Zoglauer make up the COSI Science Steering Committee.

\end{contribution}

\software{MEGAlib \citep{2006NewAR..50..629Z}, cosiburstpy \citep{cosiburstpy}, GDT-Fermi \citep{GDT-Fermi}, cosipy \citep{the_compton_spectrometer_and_imager_cosi_2025_15125776}}

\bibliography{cosi_sgrbs}{}
\bibliographystyle{aasjournalv7}

\end{document}